\documentclass[aps,prb,reprint,10pt,nofootinbib,floatfix]{revtex4-2}

\usepackage[utf8]{inputenc}
\usepackage[T1]{fontenc}
\usepackage{anyfontsize}

\usepackage{amsmath}
\usepackage{amssymb}
\usepackage{amsfonts}
\usepackage{braket}
\usepackage{mathrsfs}
\usepackage{enumitem}
\usepackage{orcidlink}

\usepackage{microtype}
\usepackage[normalem]{ulem}
\usepackage{soul}

\usepackage{graphicx}

\usepackage[dvipsnames]{xcolor}

\usepackage{etoolbox}
\makeatletter
\g@addto@macro\normalsize{%
  \abovedisplayskip 7pt plus 1.5pt minus 2pt
  \belowdisplayskip 7pt plus 1.5pt minus 2pt
  \abovedisplayshortskip 4pt plus 1pt minus 1pt
  \belowdisplayshortskip 4pt plus 1pt minus 1pt
}
\AtBeginDocument{%
  \def\@textbottom{\vskip0pt plus12pt\relax}%
  \patchcmd{\section}{0.8cm \@plus1ex \@minus .2ex}
    {14pt \@plus2pt \@minus2pt}{}%
    {\PackageError{layout}{Section spacing patch failed}{Check the REVTeX version.}}%
  \patchcmd{\section}{0.5cm}{7pt}{}%
    {\PackageError{layout}{Section after-skip patch failed}{Check the REVTeX version.}}%
  \patchcmd{\subsection}{.8cm \@plus1ex \@minus .2ex}
    {11pt \@plus1.5pt \@minus1.5pt}{}%
    {\PackageError{layout}{Subsection spacing patch failed}{Check the REVTeX version.}}%
  \patchcmd{\subsection}{.5cm}{6pt}{}%
    {\PackageError{layout}{Subsection after-skip patch failed}{Check the REVTeX version.}}%
}
\makeatother

\usepackage{hyperref}
\hypersetup{
    colorlinks=true,
    citecolor=Purple,
    linkcolor=Purple,
    urlcolor=Purple,
    linktocpage=true,
    breaklinks=true
}

\usepackage[capitalize]{cleveref}

\begin{document}

\title{Dynamical Schwinger Production of Dipole--Antidipole Pairs in a Fractonic Lattice}

\author{Celio R. Muniz}
\email{celio.muniz@uece.br}
\affiliation{Universidade Estadual do Cear\'a (UECE),
Faculdade de Educa\c{c}\~ao, Ci\^encias e Letras de Iguatu,
Av. D\'ario Rabelo s/n, Iguatu -- CE, 63500-000, Brazil}

\author{R. N. Costa Filho}
\email{rai@fisica.ufc.br}
\affiliation{Departamento de F\'isica, Universidade Federal do Cear\'a,
Caixa Postal 6030, Campus do Pici, 60455-760 Fortaleza,
Cear\'a, Brazil}

\begin{abstract}
We investigate field-induced excitation production from the unit-filled Mott background of a one-dimensional dipole-conserving Bose--Hubbard chain driven by a periodic quadratic potential, realizing a time-dependent rank-two electric field. Unlike protocols that manipulate pre-existing dipolar or fractonic excitations, we address their production from an initially Mott-like state without prepared dipolar or fractonic excitations. A single correlated hop directly nucleates the compact $|030\rangle$ configuration at the atomic energy $3U$, while a lower manifold derived from $2U$ contains spatially resolved dipole--antidipole configurations. Exact diagonalization shows that finite hopping turns this lower manifold into a dispersive band whose finite-chain edge lies substantially below the compact $3U$ scale. Despite this lower energetic edge, its normalized odd-harmonic spectral weight is perturbatively suppressed as $(J/U)^2$, reflecting virtual hybridization of the initial and final eigenstates. A consistent second-order treatment of the $|1111\rangle\leftrightarrow|0220\rangle$ channel includes diagonal self-energy shifts and predicts a weak-field resonance displacement confirmed by exact time evolution. Extended-chain dynamics shows genuine dipole--antidipole separation together with redistribution into higher-energy many-body sectors. The resulting mechanism provides a condensed-matter analogue of multiphoton dynamical Schwinger production in which exact dipole conservation replaces mobile opposite charges by mobile opposite dipoles, while isolated fractonic charges remain immobile.
\end{abstract}

\maketitle

\section{Introduction}

The Schwinger effect is a canonical example of field-induced production out of a gapped vacuum. In relativistic quantum electrodynamics, an electric field can promote virtual charged fluctuations into real particle--antiparticle pairs. For time-dependent fields this includes a multiphoton regime in which several drive quanta combine to bridge the excitation gap \cite{Schwinger1951,DunneSchubert2005,GelisTanji2016}. The same organizing idea has proved useful in condensed-matter and cold-atom systems: one asks which excitations can be created from a gapped reference state, what energetic threshold must be crossed, what microscopic matrix element couples the reference state to the excited sector, and what real-time dynamics follows after production. In particular, Schwinger-type production and dielectric breakdown have been formulated for Mott systems, including nonlinear and multiphoton doublon production \cite{GreenSondhi2005,OkaAoki2010,Oka2012}. Tilted Bose--Hubbard Mott insulators and ultracold-atom simulators provide closely related realizations of this viewpoint \cite{SachdevSenguptaGirvin2002,Queisser2012,KolodrubetzEtAl2012,Kasper2016}, while cold-atom platforms more broadly provide a controlled setting for lattice gauge-theory simulation \cite{AidelsburgerEtAl2022}.

Dipole-conserving systems alter the problem at the level of kinematics.  Besides total particle number, the dipole moment is exactly conserved.  Translating an isolated charge-like excitation by one lattice site changes the total dipole moment, which is the elementary origin of fractonic immobility in scalar-charge-type settings \cite{Pretko2017,Pretko2018,PretkoReview2020,NandkishoreHermele2019}.  A neutral \emph{dipole}, however, can be mobile because rigid translation of the composite preserves its intrinsic dipole moment.  The coexistence of immobile isolated charges and mobile dipolar composites is therefore a defining consequence of fractonic kinematics.  The compatibility of mobile dipolar composites with exact dipole conservation is well established in fractonic dynamics \cite{PaiPretkoNandkishore2019}; here an oppositely oriented dipole pair is instead produced resonantly from a Mott background by a periodic tensor electric field.

This distinction determines what a Schwinger-type production problem can mean in the present setting.  The analogue of $e^-e^+$ creation is not a pair of freely mobile isolated fracton charges.  Exact dipole conservation instead selects composite configurations whose total excess charge and total dipole both vanish.  The physically relevant questions are therefore: which dipolar pair can be nucleated from the Mott background, where the lowest finite-chain spectral edge lies in the Mott-connected fragment, how strongly the microscopic drive couples to that sector, and whether the produced pair develops genuine spatial separation.

Dipole-conserving Bose--Hubbard chains provide an explicit setting for these questions. Their constrained dynamics, phase structure, fragmentation properties, and relation to strongly tilted optical lattices have been developed in
Refs.~\cite{LakeHermeleSenthil2022,LakeEtAl2023,BoeslEtAl2024,
OhEtAl2024,KimEtAl2025,Liu2025Fractonic}.
A further key ingredient is the ability to couple such systems to synthetic tensor gauge fields. A quadratic spatial potential is the lattice analogue of a rank-two scalar potential; its second derivative is spatially uniform and plays the role of a tensor electric field \cite{ZhangLvZhou2025,ZhangXuZhouZhang2026}.
Zhang and Zhang recently considered the periodically modulated version of this field and derived the associated rotating-frame photon-assisted correlated hopping, using it to manipulate \emph{prepared} dipolar and fractonic excitations \cite{ZhangZhang2026}. Building on this microscopic framework, we address a distinct problem:
field-induced production from the Mott background. We determine which multipolar sector is created, how its lowest spectral edge differs from the strongest microscopic nucleation channel, and whether the resulting dipolar objects subsequently separate in real space.

Concretely, the initial state is the strongly coupled, unit-filled Mott background with no prepared excitation, and the question is what the oscillating tensor field \emph{produces}.  A single action of the correlated hopping operator gives
\begin{equation}
|111\rangle\longrightarrow|030\rangle,
\end{equation}
with excess occupation $(-1,+2,-1)$, vanishing excess charge and dipole moment, and atomic-limit interaction energy $3U$.  Locally, this channel has the largest direct matrix element and supports a clean multiphoton Floquet resonance $n\hbar\omega\simeq3U$.  The corresponding Bessel-dressed coupling is $\sqrt{6}\,J\,\mathcal{J}_n(A/\hbar\omega)$.

The extended-chain physics shows, however, that the compact $3U$ configuration is not the lowest pair sector.  The identity
\begin{equation}
(-1,+2,-1)=(-1,+1)+(+1,-1)
\end{equation}
reveals the compact object as the minimum-separation decomposition of two oppositely oriented dipoles.  Spatially resolved states with two doublons and two holes instead cost $2U$ in the atomic limit.  The elementary family $|02\,1\cdots1\,20\rangle$ corresponds to nearest-neighbor dipoles, but the complete $2U$ manifold is slightly richer: it also contains states in which each hole--doublon dipole has a larger internal extent.  This refinement matters because the distance between the doublons is not, in general, the same as the distance between the dipole centers.  It does not remove the pair interpretation; it identifies more precisely the internal and relative coordinates of the produced multipolar objects.

Finite $J$ hybridizes the atomic $2U$ degeneracy into a dispersive many-body band.  Consequently three physically distinct quantities must be kept separate throughout the paper:
\begin{equation}
\begin{aligned}
&\text{lowest finite-chain spectral edge},\\
&\text{microscopic production matrix element},\\
&\text{real-space dynamics of the produced sector}.
\end{aligned}
\end{equation}
The lowest finite-chain spectral edge belongs to the $2U$-derived band; the strongest direct local coupling points to the compact $3U$ configuration; and the separating mobile objects are dipolar excitations rather than isolated fracton charges.

A second distinction concerns the response operator.  In the strict atomic Mott state, one action of $\mathcal O=\sum_jO_j$ creates only compact $030$ states, so a spectral function built from $\mathcal O$ is a convenient diagnostic.  For the dressed finite-$J$ ground state, however, the physical one-quantum response is generated by the quadratic-potential operator $Q_2=\sum_jj^2n_j$, or equivalently by the odd Floquet harmonic $\mathcal O-\mathcal O^\dagger$.  We therefore formulate the weak-field golden-rule response in the laboratory frame and use the exact commutator between $Q_2$ and $H_0$ to connect it to the rotating-frame operator.  Numerically this correction is small for the parameters used in the spectral plots, but it is required for a frame-consistent interpretation.

The local second-order route to $|0220\rangle$ also requires one refinement.  Eliminating the intermediate $|0301\rangle$ and $|1030\rangle$ states generates not only the off-diagonal coupling $G_m=\mathcal{O}(J^2/U)$ but also diagonal self-energy shifts of the same perturbative order.  The effective resonance is therefore not generically the bare condition $m\hbar\omega=2U$.  In the weak-field regime the shift can be larger than the effective Rabi width, whereas for the representative $\alpha=1$ benchmark the initial- and final-state shifts nearly cancel.  This explains why the bare-resonance approximation remains accurate for the representative strong-drive benchmark while failing sharply in a discriminating weak-field test.

Finally, the extended-chain quantity $P_{2U}(t)$ must be interpreted with care.  It is the probability that the \emph{entire chain} lies in the atomic $2U$ manifold, not a density of all produced excitations.  As the chain grows, significant probability is transferred to sectors of atomic energy $4U$ and above.  A decrease of the finite-window maximum of $P_{2U}$ therefore does not by itself signal suppressed production; it also reflects the opening of more complex many-body channels.

The paper is organized accordingly.  Section~\ref{sec:model} introduces the microscopic chain and the tensor drive.  Section~\ref{sec:channel} derives compact nucleation and then resolves the geometry of the $2U$ dipole--antidipole manifold.  Section~\ref{sec:rotating} develops the rotating-frame Floquet description.  Section~\ref{sec:production} derives the local multiphoton dynamics and the frame-consistent many-body spectral response.  Section~\ref{sec:hierarchy} derives the second-order $0220$ amplitude together with its diagonal self-energy shifts.  Section~\ref{sec:EDspectrum} presents the exact-diagonalization spectrum, finite-chain lower spectral edge, spectral-weight scaling, collective enhancement, and Hilbert-space fragmentation.  Section~\ref{sec:exactvalidation} benchmarks the local effective laws against exact time evolution, including the weak-field shifted resonance.  Section~\ref{sec:pairdynamics} follows the extended-chain dynamics, separating the two-hole/two-doublon $2U$ population, transfer to higher sectors, and the geometric separation of the elementary $r=1$ dipole--antidipole component.  Section~\ref{sec:discussion} summarizes the dynamical Schwinger dictionary and experimental observables.

\section{Dipole-conserving Bose--Hubbard chain and tensor drive}
\label{sec:model}

We consider the one-dimensional dipole-conserving Bose--Hubbard model used in Ref.~\cite{ZhangZhang2026},
\begin{equation}
\hat H(t)=\hat H_0+\hat H_e(t),
\label{eq:Htotal}
\end{equation}
with
\begin{equation}
\hat H_0=-J\sum_j\left(\hat O_j+\hat O_j^\dagger\right)
+\frac{U}{2}\sum_j\hat n_j(\hat n_j-1),
\label{eq:H0}
\end{equation}
where
\begin{equation}
\hat O_j\equiv \hat b_{j-1}\hat b_j^{\dagger 2}\hat b_{j+1},
\label{eq:Oj}
\end{equation}
and
\begin{equation}
\hat H_e(t)=\frac{A}{2}\cos(\omega t)\sum_j j^2\hat n_j.
\label{eq:He}
\end{equation}
Here $\hat b_j$ and $\hat b_j^\dagger$ are bosonic annihilation and creation operators, $\hat n_j=\hat b_j^\dagger\hat b_j$, $J$ is the correlated tunneling amplitude, $U>0$ is the on-site repulsion, and $A$ and $\omega$ are the amplitude and angular frequency of the quadratic drive. We take the lattice spacing as the unit of length unless stated otherwise.

The unusual kinetic term in Eq.~\eqref{eq:H0} is central. The operator $\hat O_j$ removes one boson from each neighboring site $j\! -\!1$ and $j\!+\!1$ and creates two bosons on the central site $j$. Its Hermitian conjugate implements the reverse process. Thus particles do not hop independently: motion occurs only through correlated rearrangements.

Define total particle number and dipole moment,
\begin{equation}
\hat N=\sum_j\hat n_j,
\qquad
\hat P=\sum_j j\hat n_j.
\label{eq:NP}
\end{equation}
For a single action of $\hat O_j$, the occupation changes are
\begin{equation}
(\Delta n_{j-1},\Delta n_j,\Delta n_{j+1})=(-1,+2,-1).
\label{eq:localchange}
\end{equation}
Therefore
\begin{equation}
\Delta N=(-1)+2+(-1)=0,
\label{eq:DeltaN}
\end{equation}
and
\begin{align}
\Delta P
&=-(j-1)+2j-(j+1)
=0.
\label{eq:DeltaP}
\end{align}
Consequently,
\begin{equation}
[\hat H_0,\hat N]=[\hat H_0,\hat P]=0.
\label{eq:conservationH0}
\end{equation}
The drive in Eq.~\eqref{eq:He} is diagonal in the occupation basis, so it also commutes with both $\hat N$ and $\hat P$,
\begin{equation}
[\hat H_e(t),\hat N]=[\hat H_e(t),\hat P]=0.
\label{eq:conservationdrive}
\end{equation}
Hence the full driven evolution preserves particle number and dipole moment exactly.

A useful microscopic identity makes the meaning of the quadratic drive particularly transparent.  Under one action of $\hat O_j$, the second moment changes by
\begin{equation}
\Delta Q_2=2j^2-(j-1)^2-(j+1)^2=-2.
\label{eq:uniformphaseidentity}
\end{equation}
The result is independent of $j$.  Therefore every allowed correlated hop acquires the same drive-induced phase in the rotating frame.  This is the microscopic reason why a quadratic potential acts as a spatially uniform rank-two electric field for the present hopping process: there is no residual position-dependent Peierls phase after the transformation.  A related consequence concerns rigid translations of a produced neutral pair.  Under a translation by $s$ sites, the excess second moment transforms as $\delta Q_2\to\delta Q_2+2s\,\delta P+s^2\delta N$.  Because the produced dipole--antidipole sector has $\delta N=\delta P=0$, a rigid translation of the entire pair leaves $\delta Q_2$ unchanged.  The tensor field therefore does no work on the center-of-mass translation of the neutral pair and does not generate conventional center-of-mass Bloch oscillations; it couples instead to changes of the internal and relative coordinates, as made explicit below in Eq.~\eqref{eq:linearSeparationEnergy}.

The quadratic potential in Eq.~\eqref{eq:He} is the lattice realization of a time-dependent rank-two electric field. Writing $x=ja$ for lattice spacing $a$, the temporal tensor potential may be identified, up to convention-dependent factors, with a quadratic scalar potential $A_0\propto x^2\cos\omega t$. Its second spatial derivative is uniform, so the corresponding tensor electric field $E_{xx}(t)$ is spatially homogeneous and oscillatory \cite{ZhangLvZhou2025,ZhangZhang2026}. This is the higher-rank analogue of driving an ordinary charged system with a spatially uniform electric field.

At unit filling and in the strongly interacting regime
\begin{equation}
J/U\ll1,
\label{eq:strongcoupling}
\end{equation}
the zeroth-order ground state is the Mott product state
\begin{equation}
\ket{\mathrm{MI}}=\prod_j \hat b_j^\dagger\ket{0}
=\ket{\ldots111111\ldots}.
\label{eq:MI}
\end{equation}
For finite but small $J/U$, the exact ground state contains virtual admixtures generated by the correlated hopping. Throughout the analytic treatment below, $\ket{\mathrm{MI}}$ is therefore understood as the leading strong-coupling reference state. This distinction will be important when interpreting the production process.

\section{Compact nucleation and dipole--antidipole quantum numbers}
\label{sec:channel}

We now identify the state generated directly from the Mott background and, crucially, distinguish its \emph{local} multipole character from the mobile degrees of freedom that emerge in the extended chain.

\subsection{Direct action of the correlated hopping operator}

Consider three consecutive unit-filled sites.  Acting with
\begin{equation}
\hat O_j=\hat b_{j-1}\hat b_j^{\dagger2}\hat b_{j+1}
\end{equation}
on the local Mott configuration gives
\begin{align}
\hat O_j|111\rangle
&=
(\hat b_{j-1}|1\rangle)\otimes
(\hat b_j^{\dagger2}|1\rangle)\otimes
(\hat b_{j+1}|1\rangle)
\nonumber\\
&=
\sqrt{1}\,\sqrt{2}\sqrt{3}\,\sqrt{1}\,
|030\rangle
\nonumber\\
&=\sqrt6\,|030\rangle .
\label{eq:030matrix}
\end{align}
The Hermitian-conjugate operator annihilates $|111\rangle$ because it contains $\hat b_j^2$ on a singly occupied site.  Hence
\begin{equation}
\langle030|\hat H_0|111\rangle=-\sqrt6\,J.
\label{eq:barematrix}
\end{equation}
The factor $\sqrt6$ is a genuine bosonic enhancement and will survive in every Floquet harmonic of the direct process.

The key point is that Eq.~\eqref{eq:030matrix} identifies the state most directly accessible from the \emph{atomic} Mott configuration.  It does not yet identify the lowest-energy state in the full connected many-body fragment.

\subsection{Charge, dipole, and second moment}

Relative to the unit-filled background, $|030\rangle$ carries
\begin{equation}
\delta\mathbf n=(-1,+2,-1).
\label{eq:pattern}
\end{equation}
The excess number and dipole vanish,
\begin{equation}
\delta N=\sum_\ell\delta n_\ell=0,
\qquad
\delta P=\sum_\ell \ell\,\delta n_\ell=0.
\end{equation}
The second moment is nonzero.  Using the position-independent identity already derived in Eq.~\eqref{eq:uniformphaseidentity},
\begin{equation}
\delta Q_2=-2.
\label{eq:Q2}
\end{equation}
Thus $030$ is a neutral, dipole-preserving local quadrupolar rearrangement.  Here the repetition of the numerical value serves only to classify the excitation; the underlying algebra has already been established in Sec.~\ref{sec:model}.

There is, however, a second and more useful decomposition:
\begin{equation}
(-1,+2,-1)=(-1,+1)+(+1,-1).
\label{eq:dipoledecomp}
\end{equation}
The two terms are oppositely oriented nearest-neighbor dipoles.  Therefore the $030$ configuration can be viewed as the \emph{minimum-separation} or compact configuration of a dipole--antidipole pair.  This language is important because the dipoles themselves may move while exact total dipole conservation is maintained.  We shall therefore avoid saying that ``fractons separate.''  The mobile objects are dipolar excitations of a fractonic, dipole-conserving medium; isolated charge-like fractons remain constrained.

\subsection{Atomic energies and the geometry of the \texorpdfstring{$2U$}{2U} manifold}

The on-site interaction energy is
\begin{equation}
E_U(n)=\frac U2n(n-1).
\end{equation}
The Mott configuration has zero interaction energy in this convention, while
\begin{equation}
E_U(030)=3U.
\end{equation}
We denote this local compact-nucleation energy by
\begin{equation}
\Delta_{030}^{(0)}=3U.
\label{eq:gap3U}
\end{equation}
It is deliberately \emph{not} called the global threshold.

To see why, consider four sites.  The configuration
\begin{equation}
|0220\rangle
\end{equation}
has excess occupations $(-1,+1,+1,-1)$ and costs only
\begin{equation}
\Delta_{0220}^{(0)}=2U.
\end{equation}
The nearest-neighbor-dipole family
\begin{equation}
|0_i\,2_{i+1}\,1\cdots1\,2_j\,0_{j+1}\rangle
\label{eq:pairfamily}
\end{equation}
contains a left dipole $(-1,+1)$ and an oppositely oriented right dipole $(+1,-1)$.  Every member contains two doublons and two holes and therefore has atomic interaction energy $2U$, independent of the distance between the two dipoles.

The full $2U$ manifold connected to the Mott state is broader than Eq.~\eqref{eq:pairfamily}.  Any state in this manifold contains two holes and two doublons, and in the connected fragments studied below their ordered positions may be written as
\begin{equation}
h_1<x_1<x_2<h_2,
\end{equation}
where $h_{1,2}$ are hole positions and $x_{1,2}$ are doublon positions.  Dipole conservation imposes equal internal sizes for the two oppositely oriented dipoles,
\begin{equation}
r=x_1-h_1=h_2-x_2.
\label{eq:rdef}
\end{equation}
It is then useful to distinguish the doublon--doublon distance
\begin{equation}
d=x_2-x_1
\label{eq:ddef}
\end{equation}
from the distance between the two dipole centers,
\begin{equation}
R=d+r.
\label{eq:Rdef}
\end{equation}
The family in Eq.~\eqref{eq:pairfamily} is the elementary $r=1$ subset.  Configurations with $r>1$, such as $|012210\rangle$, have the same atomic energy $2U$ but contain internally extended dipoles.  Thus the $2U$ manifold should be viewed as a family of $d_r\bar d_r$ pairs labelled by both an internal coordinate $r$ and an inter-dipole coordinate $R$.

The resulting physical sequence remains
\begin{equation}
030\ \longrightarrow\ 0220\ \longrightarrow\ 02\,1\cdots1\,20,
\label{eq:separationladder}
\end{equation}
but it should be read as the elementary $r=1$ branch of a broader multipolar manifold.  The compact $030$ state contains a triply occupied site and is not obtained by setting a separation parameter of the doublon family to zero.

The distinction between $r$ and $R$ is also the natural one for the tensor-field energy.  Relative to the Mott background, a general ordered $2U$ configuration has
\begin{align}
\delta Q_2
&=-h_1^2+x_1^2+x_2^2-h_2^2\nonumber\\
&=-2rR.
\label{eq:Q2generalpair}
\end{align}
Since the static quadratic potential contributes $(A/2)Q_2$, the corresponding field-energy shift is
\begin{equation}
\Delta E_E=-ArR.
\label{eq:linearSeparationEnergy}
\end{equation}
For the elementary family $r=1$, this reduces to a term linear in the dipole-center separation, $\Delta E_E=-AR$.  Equation~\eqref{eq:linearSeparationEnergy} is the higher-rank counterpart of the work term that energetically favors the separation of opposite charges in the ordinary Schwinger problem: here the field couples to the product of the internal dipole moment and the separation of the oppositely oriented dipoles.  We use this relation only to clarify the structure of the produced sector; a static-field tunneling calculation is outside the present scope.

\begin{figure*}[!t]
\centering
\includegraphics[width=0.96\linewidth,trim=5bp 31bp 45bp 54bp,clip]{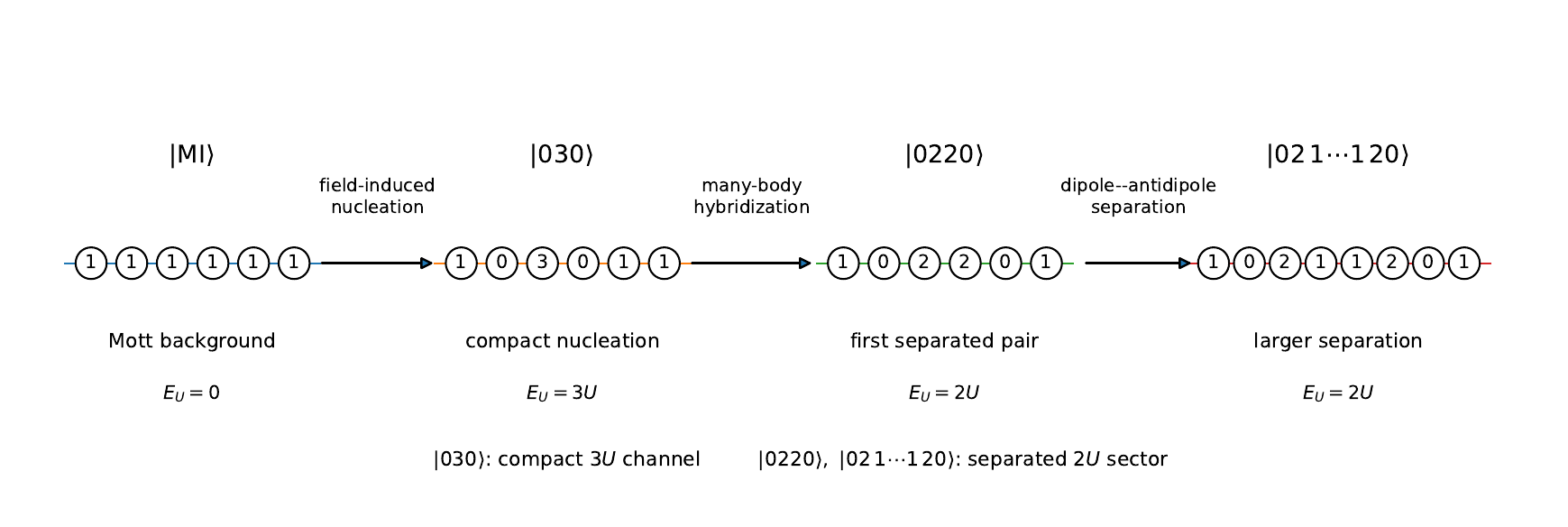}
\caption{Dipole--antidipole nucleation and the elementary separation branch.  A single correlated hop directly creates the compact $|030\rangle$ configuration from the Mott background.  Its excess pattern $(-1,+2,-1)$ decomposes into two oppositely oriented nearest-neighbor dipoles.  The $|0220\rangle$ state is the first spatially resolved $r=1$ configuration, and $|02\,1\cdots1\,20\rangle$ extends the distance between the dipole centers while preserving total particle number and total dipole moment.  The full atomic $2U$ manifold also contains $r>1$ configurations with internally extended dipoles; these are resolved explicitly in Sec.~\ref{sec:pairdynamics}.  The mobile objects are dipoles of the fractonic medium, not isolated mobile fracton charges.}
\label{fig:channels}
\end{figure*}

\section{Rotating frame and photon-assisted correlated tunneling}
\label{sec:rotating}

The quadratic drive is diagonal in the occupation basis, but it can be removed by a time-dependent unitary transformation at the price of introducing a periodic phase in the correlated hopping. This makes the multiphoton structure transparent, following the broader Floquet-engineering toolbox developed for periodically driven quantum gases \cite{Eckardt2017,BukovDAlessioPolkovnikov2015}.

We use the unitary operator \cite{ZhangZhang2026}
\begin{equation}
\hat{\mathcal U}(t)=
\exp\left[
\frac{iA}{2\hbar\omega}\sin(\omega t)
\sum_j j^2\hat n_j
\right].
\label{eq:Urot}
\end{equation}
The rotating-frame Hamiltonian is
\begin{equation}
\hat H'(t)=
\hat{\mathcal U}\hat H\hat{\mathcal U}^{-1}
+i\hbar\,\partial_t\hat{\mathcal U}\,\hat{\mathcal U}^{-1}.
\label{eq:Hrotdef}
\end{equation}
The second term exactly cancels $\hat H_e(t)$. To transform the hopping, note that
\begin{equation}
\hat{\mathcal U}\hat b_j\hat{\mathcal U}^{-1}
=e^{-if(t)j^2}\hat b_j,
\qquad
f(t)=\frac{A}{2\hbar\omega}\sin(\omega t).
\label{eq:btransform}
\end{equation}
Therefore
\begin{align}
\hat{\mathcal U}\hat O_j\hat{\mathcal U}^{-1}
&=
\exp\left\{if(t)\left[2j^2-(j-1)^2-(j+1)^2\right]\right\}\hat O_j
\nonumber\\
&=e^{-i\alpha\sin(\omega t)}\hat O_j,
\label{eq:Otransform}
\end{align}
where
\begin{equation}
{\alpha\equiv\frac{A}{\hbar\omega}.}
\label{eq:alpha}
\end{equation}
The rotating-frame Hamiltonian becomes
\begin{align}
\hat H'(t)={}&-J\sum_j\Big[
 e^{-i\alpha\sin(\omega t)}\hat O_j
+e^{+i\alpha\sin(\omega t)}\hat O_j^\dagger\Big]\notag\\
&+\frac U2\sum_j\hat n_j(\hat n_j-1).
\label{eq:Hrot}
\end{align}

The same phase can be understood directly from the multipole content of the process. Since the drive energy is $(A/2)\cos\omega t\,Q_2$, and the $111\to030$ transition changes $Q_2$ by $-2$, the instantaneous energy difference induced by the drive is
\begin{equation}
\Delta E_e(t)=-A\cos(\omega t).
\end{equation}
Integrating this energy modulation in time produces the phase $\exp[-i(A/\hbar\omega)\sin\omega t]$ in Eq.~\eqref{eq:Otransform}.

We now use the Jacobi--Anger expansion,
\begin{equation}
e^{-i\alpha\sin(\omega t)}
=
\sum_{n=-\infty}^{+\infty}
\mathcal J_n(-\alpha)e^{in\omega t},
\label{eq:JA}
\end{equation}
with $\mathcal J_n$ the Bessel function of the first kind. Since $\mathcal J_n(-\alpha)=(-1)^n\mathcal J_n(\alpha)$, the magnitude of the $n$th Fourier harmonic is $|\mathcal J_n(\alpha)|$. Hence the periodic tensor field decomposes the correlated hopping into photon-assisted channels with effective amplitudes
\begin{equation}
J_n=J\mathcal J_n(\alpha).
\label{eq:Jn}
\end{equation}
This is the microscopic Floquet mechanism by which the external field can compensate the interaction energy of a multipolar excitation.

\section{Multiphoton production of the compact dipole--antidipole configuration}
\label{sec:production}

Having identified the microscopic channel and the Floquet-dressed correlated hopping, we now derive its production dynamics. We begin with the local resonance and coherent-conversion problem, then assess the rotating-wave approximation, extend the description to the many-body spectral response, and finally make the connection with dynamical Schwinger production explicit.

\subsection{Resonance condition and local matrix element}

Let $\ket{Q_j}$ denote the normalized state obtained by replacing the local segment $111$ around site $j$ by $030$ while leaving the rest of the Mott background unchanged. In the atomic limit,
\begin{equation}
E_{Q_j}-E_{\mathrm{MI}}=3U.
\end{equation}
The $n$th Floquet harmonic transfers energy $n\hbar\omega$. Therefore the resonance condition is
\begin{equation}
{n\hbar\omega\simeq3U.}
\label{eq:resonance}
\end{equation}
For example, the one-, two-, and three-quantum resonances occur at
\begin{equation}
\hbar\omega\simeq3U,\qquad
\hbar\omega\simeq\frac{3U}{2},\qquad
\hbar\omega\simeq U,
\end{equation}
respectively.

Using Eq.~\eqref{eq:030matrix}, the local matrix element associated with the $n$th harmonic is
\begin{equation}
{
|g_n|
=\sqrt6\,J\left|\mathcal J_n\left(\frac{A}{\hbar\omega}\right)\right|.
}
\label{eq:gn}
\end{equation}
The factor $\sqrt6$ is purely bosonic and follows from the enhancement associated with creating two additional bosons on a site already occupied by one boson.

In the weak-drive limit $\alpha\ll1$, the leading Bessel functions behave as
\begin{equation}
\mathcal J_n(\alpha)\simeq\frac1{n!}\left(\frac{\alpha}{2}\right)^n,
\qquad n\ge0.
\label{eq:Besselweak}
\end{equation}
Consequently,
\begin{equation}
|g_n|\propto J\left(\frac{A}{\hbar\omega}\right)^n.
\end{equation}
This is the expected hierarchy for an $n$-quantum absorption process. In particular,
\begin{equation}
|g_1|\simeq\sqrt6\,J\frac{A}{2\hbar\omega}.
\label{eq:g1weak}
\end{equation}
At the one-quantum resonance $\hbar\omega\simeq3U$, this becomes $|g_1|\simeq\sqrt6 JA/(6U)$. This weak-drive scaling is consistent with laboratory-frame time-dependent perturbation theory: retaining the $\mathcal{O}(J/U)$ virtual admixture of $\ket{030}$ in the strong-coupling eigenstates makes the diagonal drive generate an off-diagonal matrix element of order $\sqrt6 JA/(3U)\cos\omega t$, whose resonant Fourier component is $\sqrt6 JA/(6U)$. We use this only as a consistency check on the scaling, not as an independent derivation of the full rotating-frame result.

\subsection{Coherent conversion in an isolated local channel}

To expose the basic resonance physics, first consider a local three-site sector, or equivalently an isolated avoided crossing in Floquet space, in which the relevant states are $\ket{111}$ and $\ket{030}$. Near the $n$th resonance, a rotating-wave approximation gives the effective two-level Hamiltonian
\begin{equation}
H_{\mathrm{eff}}^{(n)}=
\begin{pmatrix}
0 & g_n\\
 g_n^* & \delta_n
\end{pmatrix},
\qquad
\delta_n\equiv3U-n\hbar\omega.
\label{eq:Heff2}
\end{equation}
If the system starts in $\ket{111}$, the probability of occupying $\ket{030}$ at time $t$ is
\begin{equation}
{
P_Q^{(n)}(t)=
\frac{4|g_n|^2}{\delta_n^2+4|g_n|^2}
\sin^2\left[
\frac{t}{2\hbar}
\sqrt{\delta_n^2+4|g_n|^2}
\right].
}
\label{eq:Rabi}
\end{equation}
Exactly on resonance,
\begin{equation}
{
P_Q^{(n)}(t)=
\sin^2\left[
\frac{\sqrt6 J}{\hbar}
\mathcal J_n\left(\frac{A}{\hbar\omega}\right)t
\right].
}
\label{eq:resonantRabi}
\end{equation}
Equation~\eqref{eq:resonantRabi} describes coherent creation and recombination of the local multipolar excitation. It should not be interpreted as irreversible vacuum decay. In a finite isolated two-level system, population oscillates back and forth between the two configurations.

This distinction is conceptually important. The elementary lattice problem
is closer to a driven avoided crossing than to an infinite-volume decay
process. In an extended many-body spectrum, dispersion and multiple final
states instead motivate a golden-rule spectral description of the initial
production response.

\subsection{Quantitative validity of the rotating-wave approximation}
\label{subsec:RWAvalidity}

The rotating-wave approximation is controlled by the ratio between each discarded
Floquet coupling and its own detuning, rather than by the relative sizes of the
Bessel functions alone.  Near the one-quantum $|111\rangle\leftrightarrow|030\rangle$
resonance, the $\ell$th harmonic has coupling
\begin{equation}
g_\ell=\sqrt6\,J\,\mathcal J_\ell(\alpha)
\end{equation}
and detuning $3U-\ell\hbar\omega$.  A useful dimensionless control parameter is therefore
\begin{equation}
\epsilon_\ell^{(030)}
=
\frac{|g_\ell|}{|3U-\ell\hbar\omega|},
\qquad \ell\neq1.
\label{eq:RWAparameter}
\end{equation}
For $\alpha=1$ at the nominal one-quantum resonance $\hbar\omega=3U$, the largest
discarded contributions are
\begin{align}
\epsilon_0^{(030)}&\simeq3.12\times10^{-2},&
\epsilon_{-1}^{(030)}&\simeq8.98\times10^{-3},\\
\epsilon_2^{(030)}&\simeq4.69\times10^{-3}.&&
\end{align}
for $J/U=0.05$.  They scale linearly with $J/U$, so even at $J/U=0.10$ the
largest ratio remains about $6.2\times10^{-2}$.  Thus the nonresonant harmonics
are perturbatively controlled in the strong-coupling regime used below.

The leading effect of the discarded harmonics is a second-order
ac-Stark/Bloch--Siegert-type shift. Since the off-resonant harmonics shift
the two levels in opposite directions, the shift of their relative
separation contains the sum of the two level shifts, giving the factor
of two below. Parametrically,
\begin{equation}
\delta_{\rm BS}^{(030)}
\simeq
2\sum_{\ell\neq1}
\frac{6J^2\mathcal J_\ell^2(\alpha)}
{3U-\ell\hbar\omega}
=\mathcal{O}(J^2/U).
\label{eq:BSshift}
\end{equation}
At $\alpha=1$ and $\hbar\omega=3U$ we find
$\delta_{\rm BS}^{(030)}/U\simeq1.08\times10^{-3}$, $6.73\times10^{-3}$,
and $2.69\times10^{-2}$ for $J/U=0.02$, $0.05$, and $0.10$, respectively.
The RWA therefore reproduces the leading resonant dynamics accurately for
$J/U\lesssim0.05$, while small frequency shifts can accumulate a visible phase
error over many Rabi cycles.  The direct comparison with the full time-dependent Hamiltonian in Sec.~\ref{sec:exactvalidation} quantifies this statement.

A separate approximation must be distinguished from this Floquet-harmonic test.  The ratios in Eq.~\eqref{eq:RWAparameter} control nonresonant harmonics \emph{inside the local $111\leftrightarrow030$ truncation}; they do not test leakage from that local subspace into other states of the extended many-body fragment.  The latter is a Hilbert-space truncation issue, and it is quantified independently by the exact-diagonalization spectral weights in Sec.~\ref{sec:EDspectrum}.  A small $\epsilon_\ell^{(030)}$ therefore guarantees neither negligible $2U$-band weight nor globally two-level dynamics.

Equation~\eqref{eq:Rabi} also makes the role of detuning transparent: exact resonance permits complete coherent conversion within the isolated two-level approximation, whereas finite $|\delta_n|$ lowers the maximum attainable population. For the one-quantum sector at $\alpha=1$, $|g_1|/J=\sqrt6\,\mathcal J_1(1)\simeq1.078$. The representative choices $|\delta_1|=|g_1|$ and $|\delta_1|=3|g_1|$ give maximum populations $4/5$ and $4/13$, respectively, while also changing the generalized Rabi frequency. These direct-channel curves are shown in Fig.~\ref{fig:localdynamics}(a), together with the
second-order $0220$ dynamics shown in Fig.~\ref{fig:localdynamics}(b) and derived in Sec.~\ref{sec:hierarchy}. The two panels are juxtaposed only to compare the local strong-coupling timescales; they correspond to different resonance frequencies and are not simultaneous channels of a single fixed-frequency drive. Throughout, the horizontal axis is the physical strong-coupling time variable $Jt/\hbar$.

\subsection{Production density and the physical response operator in an extended chain}

For an extended chain, finite hopping dresses the Mott state and broadens the atomic configurations into exact many-body eigenstates.  The appropriate response operator depends on which regime is being discussed.  In the laboratory frame the external perturbation is
\begin{equation}
\hat V(t)=\frac{A}{2}\cos(\omega t)\,\hat Q_2,
\qquad
\hat Q_2\equiv\sum_jj^2\hat n_j.
\label{eq:VQ2}
\end{equation}
Therefore the controlled weak-field one-quantum absorption problem is governed directly by the spectral density of $\hat Q_2$,
\begin{equation}
\mathcal S_{Q_2}(E)=
\frac1L\sum_{f\ne G}
\left|\langle f|\hat Q_2|G\rangle\right|^2
\delta(E-E_f+E_G).
\label{eq:SQ2}
\end{equation}
Because the resonant Fourier component of $(A/2)\cos\omega t$ has amplitude $A/4$, standard linear-response theory and Fermi's golden rule \cite{Mahan2000} give the one-quantum spectral production density per site
\begin{equation}
\gamma_1
=\frac{2\pi}{\hbar}\frac{A^2}{16}
\mathcal S_{Q_2}(\hbar\omega).
\label{eq:gammaQ2}
\end{equation}
Equation~\eqref{eq:gammaQ2} is a linear-response statement.  In a finite coherent spectrum it should be understood as a spectral transition formula rather than as a universal irreversible decay rate.

The same physics appears in the rotating frame.  Define
\begin{equation}
\hat{\mathcal O}=\sum_j\hat O_j.
\label{eq:Oext}
\end{equation}
With the Fourier convention $H'(t)=\sum_nH_ne^{-in\omega t}$, the $n$th hopping harmonic is proportional to
\begin{equation}
\hat X_n=\hat{\mathcal O}+(-1)^n\hat{\mathcal O}^\dagger.
\label{eq:Xn}
\end{equation}
Thus odd harmonics, including the one-quantum channel, involve
\begin{equation}
\hat X_1=\hat{\mathcal O}-\hat{\mathcal O}^\dagger,
\label{eq:X1}
\end{equation}
whereas even harmonics contain the Hermitian combination $\hat{\mathcal O}+\hat{\mathcal O}^\dagger$.  In the strict atomic Mott state $\hat{\mathcal O}^\dagger|\mathrm{MI}\rangle=0$, which is why keeping only $\hat{\mathcal O}$ reproduces the atomic-limit creation matrix element.  For the dressed finite-$J$ ground state, however, both terms must be retained.

The laboratory- and rotating-frame descriptions are connected by an exact commutator identity,
\begin{equation}
[\hat H_0,\hat Q_2]
=-2J\left(\hat{\mathcal O}-\hat{\mathcal O}^\dagger\right).
\label{eq:commutatorQ2}
\end{equation}
For an excited eigenstate $|f\rangle$ of $H_0$ with $E=E_f-E_G>0$, this gives
\begin{equation}
E\,\langle f|\hat Q_2|G\rangle
=-2J\langle f|\hat X_1|G\rangle
\end{equation}
Consequently,
\begin{equation}
\mathcal S_{Q_2}(E)
=\frac{4J^2}{E^2}\mathcal S_{X_1}(E),
\label{eq:spectralidentity}
\end{equation}
where
\begin{equation}
\mathcal S_{X_1}(E)=
\frac1L\sum_{f\ne G}
|\langle f|\hat X_1|G\rangle|^2
\delta(E-E_f+E_G).
\label{eq:SX1}
\end{equation}
In a frequency scan, fixing the physical drive amplitude $A$ is
distinct from fixing $\alpha=A/(\hbar\omega)$, since the latter requires
$A$ to vary with frequency. We verified Eq.~\eqref{eq:spectralidentity} numerically state by state for $L=10$ and $J/U=0.05$, with discrepancies at the level of numerical precision.  This identity is useful because it gives a direct physical meaning to the odd-harmonic spectral weight used below: it is the same weak-field absorption spectrum generated by the quadratic tensor potential, apart from the known kinematic factor $4J^2/E^2$.

For finite $\alpha$ the rotating-frame Hamiltonian contains several harmonics as well as the $n=0$ component $J\mathcal J_0(\alpha)$. At finite drive amplitude the Bessel decomposition is used to identify
resonant channels, while the actual dynamics is propagated with the full
time-dependent Hamiltonian in Secs.~\ref{sec:exactvalidation} and \ref{sec:pairdynamics}.

\subsection{From local Floquet nucleation to the many-body pair sector}
\label{subsec:schwingerlocal}

The formulas above establish the direct, Bessel-dressed multiphoton nucleation law of the compact $030$ configuration and identify the physical weak-field response operator.  They do not by themselves determine the lowest many-body spectral edge.  In the strict atomic Mott state,
\begin{equation}
\hat{\mathcal O}|\mathrm{MI}\rangle
=\sqrt{6}\sum_{j=2}^{L-1}|Q_j\rangle,
\label{eq:OonMI}
\end{equation}
and $\hat{\mathcal O}^\dagger|\mathrm{MI}\rangle=0$, so the one-action spectral weight is concentrated at $3U$.  For finite $J/U$, however, the exact ground state is dressed and $\hat X_1|G\rangle=(\hat{\mathcal O}-\hat{\mathcal O}^\dagger)|G\rangle$ acquires overlap with the lower $2U$-derived band.  The same overlap is visible in the physical $Q_2$ response through Eq.~\eqref{eq:spectralidentity}.

The local and extended descriptions therefore answer different questions.  The few-site Floquet problem determines the microscopic amplitude for compact nucleation and for the first separated configuration.  The exact spectrum determines which many-body eigenstates carry the response at a given energy.  The next section derives the local second-order route into $|0220\rangle$ consistently to $\mathcal{O}(J^2/U)$, including the diagonal energy shifts generated by the same virtual processes, and Sec.~\ref{sec:EDspectrum} then resolves the complete extended-chain band.

\section{From compact nucleation to the separated \texorpdfstring{$2U$}{2U} pair manifold}
\label{sec:hierarchy}

The state $|0220\rangle$ is not an unrelated competitor to $|030\rangle$.  It is the first spatially resolved member of the elementary $r=1$ branch of the dipole--antidipole manifold.  The local perturbative calculation is useful because it determines how the \emph{pure atomic Mott state} first reaches this separated configuration and, when carried consistently to second order, also determines the corresponding resonance shift.

\subsection{Two reflected second-order paths}

On four consecutive sites the two minimal paths are
\begin{align}
|1111\rangle&\rightarrow|0301\rangle\rightarrow|0220\rangle,\\
|1111\rangle&\rightarrow|1030\rangle\rightarrow|0220\rangle.
\label{eq:twoPaths}
\end{align}
For the first path,
\begin{align}
\hat O_2|1111\rangle
&=\sqrt6\,|0301\rangle,\\
\hat O_3|0301\rangle
&=\sqrt3\,\sqrt2\,|0220\rangle
=\sqrt6\,|0220\rangle.
\label{eq:bosonicSecondHop}
\end{align}
The reflected path gives the same bosonic factors.  The intermediate states are reflection partners with atomic energy $3U$, and Eq.~\eqref{eq:uniformphaseidentity} ensures that every correlated hop carries the same rotating-frame phase.  The two paths therefore add coherently, producing the numerator
\begin{equation}
2(\sqrt6J)^2=12J^2.
\label{eq:factor12}
\end{equation}

If the complete process exchanges $m$ drive quanta and the first hop exchanges $\ell$, standard second-order effective-Hamiltonian perturbation theory, applied in Floquet space \cite{BravyiDiVincenzoLoss2011}, gives the off-diagonal amplitude
\begin{equation}
G_m\equiv M_{0220}^{(2,m)}
\simeq
-12J^2\sum_{\ell\in\mathbb Z}
\frac{\mathcal J_\ell(\alpha)\mathcal J_{m-\ell}(\alpha)}
{3U-\ell\hbar\omega},
\qquad
\alpha=\frac{A}{\hbar\omega}.
\label{eq:M0220}
\end{equation}
The same elimination that generates $G_m=\mathcal{O}(J^2/U)$ also shifts the two diagonal energies.  For the local four-site truncation $|i\rangle=|1111\rangle$, $|f\rangle=|0220\rangle$, with intermediate states $|0301\rangle$ and $|1030\rangle$, each diagonal shift is obtained by summing the squared Floquet matrix elements for virtual excursions from the corresponding endpoint into these intermediate states, divided by the associated quasienergy denominators.  The resulting second-order shifts are
\begin{equation}
\Sigma_i=
-12J^2\sum_{\ell\in\mathbb Z}
\frac{\mathcal J_\ell^2(\alpha)}{3U-\ell\hbar\omega},
\label{eq:Sigmai}
\end{equation}
and
\begin{equation}
\Sigma_f=
-12J^2\sum_{\ell\in\mathbb Z}
\frac{\mathcal J_\ell^2(\alpha)}{U-\ell\hbar\omega}.
\label{eq:Sigmaf}
\end{equation}
These expressions assume that all intermediate Floquet replicas entering the
second-order elimination remain off resonance. This condition is not automatic
at the bare multiphoton resonance $m\hbar\omega=2U$. For even $m$, resonant
intermediate replicas can occur; for example, at $m=2$ one has
$\hbar\omega=U$, so that $3U-3\hbar\omega=0$ and
$U-\hbar\omega=0$. Such replicas must be retained explicitly in an enlarged
near-degenerate Floquet subspace rather than eliminated perturbatively.
The one-quantum ($m=1$) benchmarks considered below avoid this
intermediate-resonance condition.

The effective detuning is therefore
\begin{equation}
\Delta_{m,\mathrm{eff}}^{(0220)}
=2U-m\hbar\omega+\Sigma_f-\Sigma_i,
\label{eq:delta0220}
\end{equation}
so the resonance condition becomes
\begin{equation}
m\hbar\omega_{\rm res}
=2U+\Sigma_f-\Sigma_i.
\label{eq:res0220}
\end{equation}
The bare atomic condition $m\hbar\omega=2U$ is recovered only when the differential shift is negligible compared with the resonance width.

Away from intermediate resonances the local amplitude hierarchy remains
\begin{equation}
|M_{030}|\sim J,
\qquad
|G_m|\sim J^2/U.
\label{eq:hierarchyM}
\end{equation}
This is a statement about how the atomic Mott configuration enters the separated sector, not about the position of the lowest spectral edge.

\subsection{Effective coherent law and the shifted resonance}

After eliminating the intermediate states, the effective two-state Hamiltonian may be written as
\begin{equation}
H_{0220,\mathrm{eff}}^{(m)}=
\begin{pmatrix}
\Sigma_i&G_m\\
G_m^\ast&2U-m\hbar\omega+\Sigma_f
\end{pmatrix}.
\label{eq:Heff0220full}
\end{equation}
Subtracting the common energy $\Sigma_i$ gives the equivalent form
\begin{equation}
H_{0220,\mathrm{eff}}^{(m)}\sim
\begin{pmatrix}
0&G_m\\
G_m^\ast&\Delta_{m,\mathrm{eff}}^{(0220)}
\end{pmatrix}.
\label{eq:Heff0220}
\end{equation}
Starting from $|1111\rangle$ then gives
\begin{align}
P_{0220}^{(m)}(t)={}&
\frac{4|G_m|^2}{[\Delta_{m,\mathrm{eff}}^{(0220)}]^2+4|G_m|^2}\notag\\
&\times\sin^2\!\left[
\frac{t}{2\hbar}
\sqrt{[\Delta_{m,\mathrm{eff}}^{(0220)}]^2+4|G_m|^2}
\right].
\label{eq:P0220}
\end{align}
At the shifted resonance, $\Delta_{m,\mathrm{eff}}^{(0220)}=0$, this reduces to
\begin{equation}
P_{0220}^{(m)}(t)=\sin^2(|G_m|t/\hbar).
\label{eq:P0220res}
\end{equation}

The weak-field limit makes the physical importance of the diagonal terms transparent.  Near the one-quantum $2U$ resonance and for $\alpha\rightarrow0$,
\begin{equation}
\Sigma_f-\Sigma_i\longrightarrow-\frac{8J^2}{U},
\qquad
G_1\simeq-\frac{8\alpha J^2}{U}.
\label{eq:weakshift}
\end{equation}
Hence $|\Sigma_f-\Sigma_i|/|G_1|\sim1/\alpha$: reducing $J/U$ suppresses the shift and the coupling by the same power of $J$, whereas reducing the field amplitude narrows the resonance relative to its shift.  The weak-field regime therefore provides a particularly stringent test of the effective theory: the off-diagonal coupling and the differential self-energy shift are of the same order in $J$, but only the former is suppressed linearly with the drive amplitude.  Consequently, the resonance displacement becomes parametrically more important as the field is weakened.

\begin{figure*}[!t]
\centering
\includegraphics[width=1\linewidth]{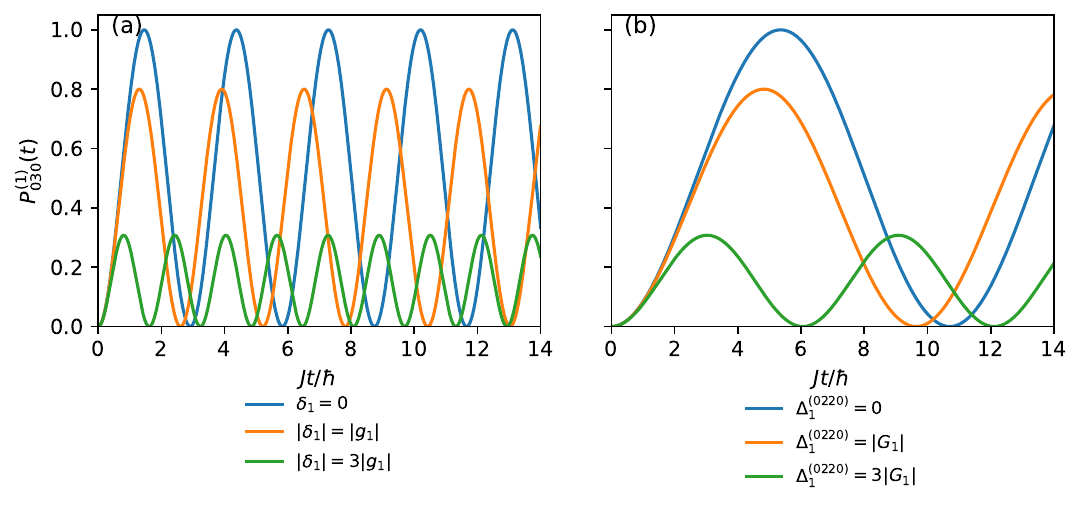}\hfill
\caption{Local strong-coupling dynamics.
(a) Direct compact $|111\rangle\leftrightarrow|030\rangle$ nucleation
in the one-quantum sector at $\alpha=1$, with resonant and detuned
Rabi curves.
(b) Effective second-order $|1111\rangle\leftrightarrow|0220\rangle$
dynamics for $J/U=0.05$ and $\alpha=1$.
The two panels refer to different one-quantum resonance neighborhoods,
near $3U$ and $2U$, respectively, and are juxtaposed only to compare
their characteristic timescales on the common $Jt/\hbar$ axis.
For panel (b), the differential shift $\Sigma_f-\Sigma_i$ is small
compared with $G_1$ at $\alpha=1$ because the two diagonal self-energies
nearly cancel, so the shifted- and bare-resonance curves are visually
almost indistinguishable; Fig.~4 shows that this approximation fails
in the weak-field limit. The longer $0220$ timescale reflects the $O(J^2/U)$ entrance amplitude
into the first separated $r=1$ configuration, not a larger excitation
gap. Here $\Delta^{(0220)}_1$ denotes the effective detuning defined in
Eq.~(\ref{eq:delta0220}).}
\label{fig:localdynamics}
\end{figure*}

For the representative strong-drive benchmark $J/U=0.05$, $\alpha=1$, and $\hbar\omega=2U$, the shifts nearly cancel: $\Sigma_i/U\simeq-0.01248$, $\Sigma_f/U\simeq-0.01364$, while $G_1/U\simeq-0.01467$.  The differential shift is only about $7.9\%$ of $|G_1|$, and the corrected resonance lies at $\hbar\omega_{\rm res}/U\simeq1.99884$.  The local curves in Fig.~\ref{fig:localdynamics} are therefore essentially unchanged on the plotted scale at $\alpha=1$; the discriminating weak-field test is presented in Fig.~\ref{fig:exactvalidation}.

The local calculation therefore explains why a lower-energy manifold can have a weaker direct entrance amplitude.  The next step is to diagonalize the actual connected chain and determine how this local perturbative structure is distributed over exact many-body eigenstates.

\section{Extended-chain spectrum, finite-size spectral edge, and spectral-weight transfer}
\label{sec:EDspectrum}

A purely local treatment cannot determine the spectral structure of the
extended chain, because neither $|030\rangle$ nor $|0220\rangle$ is an
eigenstate at finite $J$.

\subsection{Connected fragment and numerical construction}

Starting from $|\mathrm{MI}\rangle$, we repeatedly apply every allowed $\hat O_j$ and $\hat O_j^\dagger$ while keeping fixed total particle number and dipole moment.  With a local cutoff $n_{\max}=4$, the resulting connected dimensions are
\begin{equation}
\dim\mathcal H_{\rm conn}=
20,\;144,\;1192
\qquad
(L=6,8,10).
\label{eq:dimensions}
\end{equation}
Removing the $n_{\max}=4$ restriction enlarges the corresponding fragments to $22$, $167$, and $1486$ states.  The relevant strong-coupling observables are nevertheless converged: for $L=10$ and $J/U=0.05$, removing the cutoff changes $E_{\rm edge}/U$ from $1.65849$ to $1.65844$ and, with the frame-consistent odd-harmonic operator defined below, $W_{2U}$ from $0.0827110$ to $0.0827005$.  The resulting fragments remain small enough for direct diagonalization, so no tensor-network approximation is required for the system sizes used here.

Hilbert-space fragmentation is not merely a computational convenience.  For example, at $L=8$ and $n_{\max}=4$ the Mott-connected fragment has dimension $144$, whereas the full sector with the same conserved $(N,P)$ has dimension $265$.  Thus
\begin{equation}
\mathcal H_{\rm conn}(\mathrm{MI})\subsetneq\mathcal H_{N,P}.
\label{eq:fragment}
\end{equation}
The initial state selects a proper dynamical fragment even after the two obvious global conserved quantities are fixed, in the sense familiar from Hilbert-space fragmentation and shattering \cite{SalaEtAl2020,KhemaniEtAl2020}.

\subsection{The \texorpdfstring{$2U$}{2U} degeneracy becomes a dispersive band}

At $J=0$, the complete two-hole/two-doublon manifold has interaction energy $2U$; the elementary family in Eq.~\eqref{eq:pairfamily} is its $r=1$ subset.  Finite correlated hopping mixes these configurations, lifts the atomic degeneracy, and produces a dispersive band.  For $J/U=0.05$, the lowest finite-chain spectral edge decreases monotonically with system size,
\begin{equation}
\begin{aligned}
L={}&6,7,8,9,10,11,12,\\
E_{\rm edge}/U={}&1.75392,1.71167,1.68627,\\
&1.66982,1.65849,1.65031,1.64419.
\end{aligned}
\label{eq:gapsize}
\end{equation}
The decreasing increments are consistent with convergence to a lower-band edge well below $3U$.  However, fits linear in $1/L$ and $1/L^2$ over these still-small sizes give noticeably different intercepts, so we do not quote a thermodynamic-limit extrapolation.
For $L=10$ the $2U$-derived band spans approximately
\begin{equation}
1.6585U\lesssim E-E_0\lesssim2.3333U.
\label{eq:2Uband}
\end{equation}
The lower edge
\begin{equation}
E_{\rm edge}(L=10)=1.6585U
\label{eq:trueThreshold}
\end{equation}
is therefore the lowest excitation energy in the Mott-connected finite-chain spectrum for these parameters.  We refer to this quantity as a finite-chain pair-sector edge rather than as a thermodynamic threshold.  The compact $3U$ object is not the lowest excitation; its importance comes instead from its large direct microscopic matrix element.

The $3U$-derived weight behaves very differently: at $L=10$, $J/U=0.05$ the prominent weight is confined to the narrow interval
\begin{equation}
3.02125U\lesssim E-E_0\lesssim3.10513U.
\end{equation}
This contrast provides a spectral distinction between a quasi-local compact manifold near $3U$ and a much more dispersive dipole--antidipole band derived from $2U$.

\subsection{Spectral weight of the physical odd-harmonic response}

We now resolve the weak-field response over the exact finite-$J$ spectrum.  The physically relevant one-quantum rotating-frame operator is
\begin{equation}
\hat X_1=\hat{\mathcal O}-\hat{\mathcal O}^\dagger,
\end{equation}
as derived in Sec.~\ref{sec:production}.  In the strict atomic Mott limit
$\hat{\mathcal O}^\dagger|\mathrm{MI}\rangle=0$, so
$\hat X_1|\mathrm{MI}\rangle=\hat{\mathcal O}|\mathrm{MI}\rangle$,
with the latter given by Eq.~\eqref{eq:OonMI}. Hence
\begin{equation}
\|\hat X_1|\mathrm{MI}\rangle\|^2=6(L-2).
\label{eq:sumrule}
\end{equation}
Thus the corrected operator reproduces the same atomic $3U$ line and bosonic sum rule while remaining valid for the dressed ground state.

\begin{figure*}[!t]
\centering
\includegraphics[width=1\linewidth]{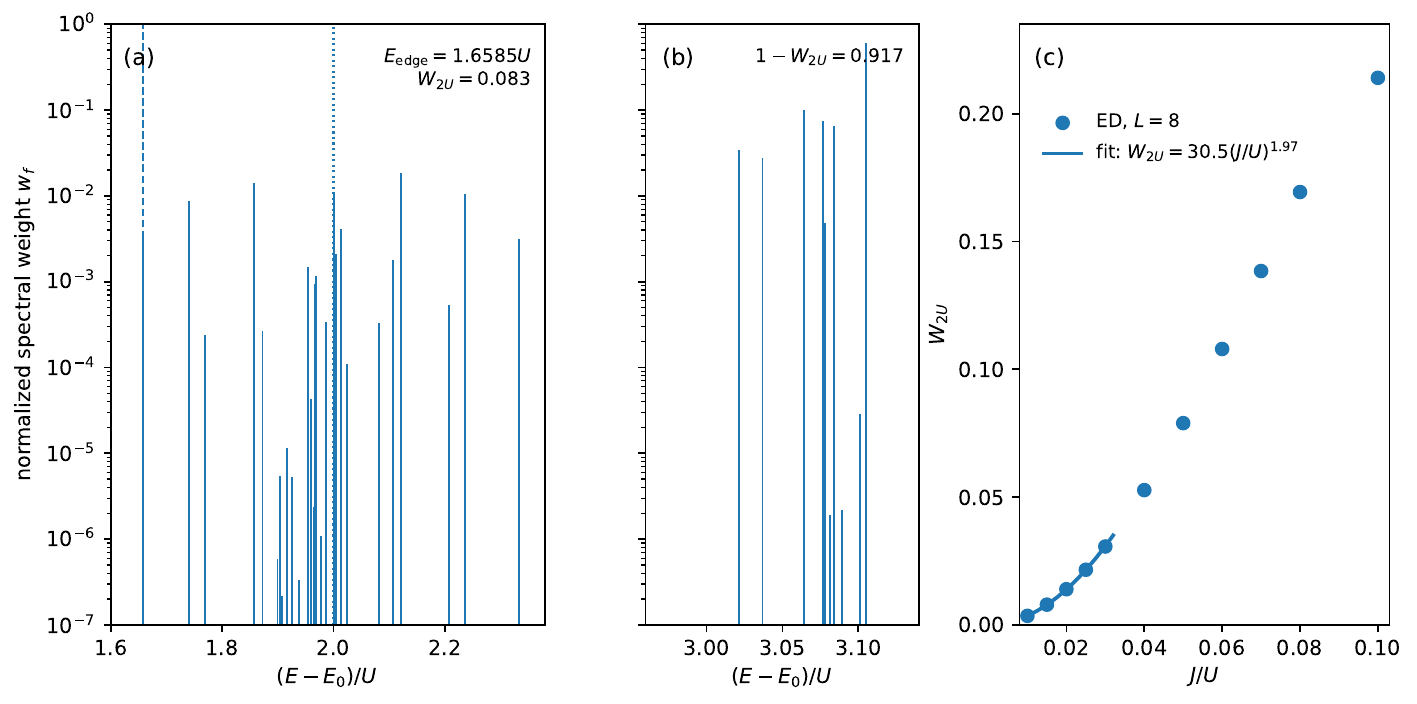}
\caption{Extended-chain spectral structure from exact diagonalization
using the frame-consistent odd-harmonic operator
$\hat X_1=\hat O-\hat O^\dagger$.
Panel (a) shows the normalized state-by-state spectral weight in the
broad $2U$-derived band for $L=10$ and $J/U=0.05$, with finite-chain
lower edge $E_{\rm edge}(L=10)=1.6585U$ and integrated weight
$W_{2U}=0.082711$. The dashed line marks $E_{\rm edge}(L=10)$,
while the dotted line at $E-E_0=2U$ indicates the corresponding
atomic-limit pair-production threshold. Panel (b) resolves the narrow $3U$-derived structure carrying the complementary plotted weight $1-W_{2U}=0.917$; this quantity is the
complement of the integrated lower-band weight rather than a separately
defined $3U$-window integral. Panel (c) shows $W_{2U}$ for $L=8$ as a function of $J/U$; the strong-coupling fit $30.5(J/U)^{1.97}$ confirms the expected quadratic scaling. Through Eq.~\eqref{eq:spectralidentity}, the same spectral information determines the weak-field response to the physical quadratic-potential operator $\hat Q_2$.}
\label{fig:spectrum}
\end{figure*}

For the plotted discrete spectrum we normalize the excited-state weights as
\begin{equation}
w_f=
\frac{|\langle f|\hat X_1|G\rangle|^2}
{\sum_{f\ne G}|\langle f|\hat X_1|G\rangle|^2}.
\label{eq:wfnorm}
\end{equation}
The integrated lower-band weight is defined by
\begin{equation}
W_{2U}=\sum_{1<(E_f-E_G)/U<2.5}w_f.
\label{eq:W2def}
\end{equation}
For $L=10$ and $J/U=0.05$ we obtain
\begin{equation}
W_{2U}=0.082711,
\qquad
1-W_{2U}=0.917289.
\label{eq:W2number}
\end{equation}
The state-resolved spectra displayed in Figs.~\ref{fig:spectrum}(a) and \ref{fig:spectrum}(b) show that the complementary weight, $1-W_{2U}=0.917289$, is overwhelmingly concentrated in the narrow $3U$-derived manifold.  We therefore report the directly integrated lower-band weight $W_{2U}$ together with its complement, without introducing a separate independently integrated $W_{3U}$ quantity. For comparison, using $\hat{\mathcal O}$ alone gives $W_{2U}=0.082495$.  The numerical difference is only about $0.26\%$, showing that the spectral picture is quantitatively stable, while Eq.~\eqref{eq:wfnorm} provides the frame-consistent odd-harmonic definition.

It is important to distinguish this normalized odd-harmonic weight from the
corresponding normalized weight of the physical quadratic-potential operator
$\hat Q_2$. Equation~(69) relates the unnormalized spectra through the
energy-dependent factor $4J^2/E^2$, so their separately normalized integrated
fractions need not coincide. For $L=10$, $J/U=0.05$, and $n_{\max}=4$, the
lower-band fraction is $8.2711\%$ for $\hat X_1$, as reported above, whereas
the separately normalized $\hat Q_2$ spectrum gives $17.9982\%$ over the same
energy window.

The lower band has the smallest excitation energy, but most of the normalized
odd-harmonic spectral weight remains near the compact $3U$ manifold. This is
not a contradiction because spectral edge and spectral accessibility are
distinct. At finite $J/U$, virtual admixtures in both the initial Mott-like
ground state and the final eigenstates generate a transition amplitude into
the lower band at first order in $J/U$ relative to the direct compact process.
The corresponding spectral \emph{weight} must therefore scale as $(J/U)^2$.

Exact diagonalization confirms this expectation, as shown in
Fig.~\ref{fig:spectrum}(c).  For $L=8$ and $J/U\le0.03$,
\begin{equation}
W_{2U}\simeq30.5\left(\frac JU\right)^{1.97},
\label{eq:W2fit}
\end{equation}
fully consistent with
\begin{equation}
W_{2U}\propto(J/U)^2.
\label{eq:W2scale}
\end{equation}
The prefactor is specific to the finite-$L$ normalization and the chosen
energy window; the robust result is the near-quadratic exponent. Through
Eq.~(\ref{eq:spectralidentity}), the spectral information in Figs.~3(a) and 3(b) may be read
either as the spectral redistribution of the odd Floquet harmonic or,
after the known $E^{-2}$ factor, as the physical weak-field response to
the quadratic potential.

\subsection{Collective bright-state enhancement and its limitation}

The direct compact states also carry a simple collective enhancement.
Using Eq.~\eqref{eq:OonMI}, introduce the normalized bright superposition
\begin{equation}
|B\rangle=\frac1{\sqrt{L-2}}\sum_{j=2}^{L-1}|Q_j\rangle,
\end{equation}
the $n$th Floquet harmonic couples with
\begin{equation}
g_{\rm coll}^{(n)}
=
\sqrt{L-2}\,\sqrt6J\mathcal J_n(\alpha).
\label{eq:gcoll}
\end{equation}
This $\sqrt{L-2}$ factor explains the enhanced \emph{initial} coherent response of the translationally summed operator.  It must not be interpreted as instantaneous conversion in the thermodynamic limit.  As the spectrum becomes denser, the single bright-state Rabi picture crosses over to many-state hybridization and a spectral/rate description.

Dynamically, the narrow $3U$ manifold in a finite chain naturally exhibits
coherent beating, whereas the broad $2U$ band is the more natural setting
for a continuum-like production density.

\section{Microscopic benchmark of the local effective laws}
\label{sec:exactvalidation}

We now benchmark the local effective descriptions against exact propagation of the full time-dependent problem.  The propagation is performed with the full periodic hopping phases, not with a rotating-wave Hamiltonian, and the four-site $0220$ benchmark retains the intermediate $|0301\rangle$ and $|1030\rangle$ states explicitly.  We set $U=\hbar=1$ and first take $\alpha=1$.

For the direct $|111\rangle\leftrightarrow|030\rangle$ channel, the three-site sector is driven at the nominal one-quantum resonance $\hbar\omega=3U$.  Over $0\le Jt/\hbar\le14$, the maximum deviation from the isolated-channel expression in Eq.~\eqref{eq:Rabi} is approximately $0.020$, $0.070$, and $0.201$ for $J/U=0.02$, $0.05$, and $0.10$, respectively.  At $J/U=0.05$ the first maximum remains very close to the analytic prediction and reaches essentially complete compact-state conversion.  The growing long-time deviation is mainly due to micromotion and higher-order frequency shifts rather than a failure of the leading $\mathcal{O}(J)$ matrix element.

For the second-order $|1111\rangle\leftrightarrow|0220\rangle$ channel, the comparison must use the shifted detuning in Eq.~\eqref{eq:delta0220}.  At $\alpha=1$ and $\hbar\omega=2U$ the diagonal shifts nearly cancel, so the corrected effective curve remains close to the bare-resonance result on the scale of the upper panel.  The maximum deviations over the same time window are approximately $0.015$, $0.168$, and $0.776$ for $J/U=0.02$, $0.05$, and $0.10$.  The deterioration with increasing $J/U$ reflects both higher-order phase corrections and the growing transient occupation of the triply occupied intermediate states; the four-site two-level elimination is therefore a controlled strong-coupling description, not a uniformly accurate long-time theory.

\begin{figure*}[!t]
\centering
\includegraphics[width=1\linewidth]{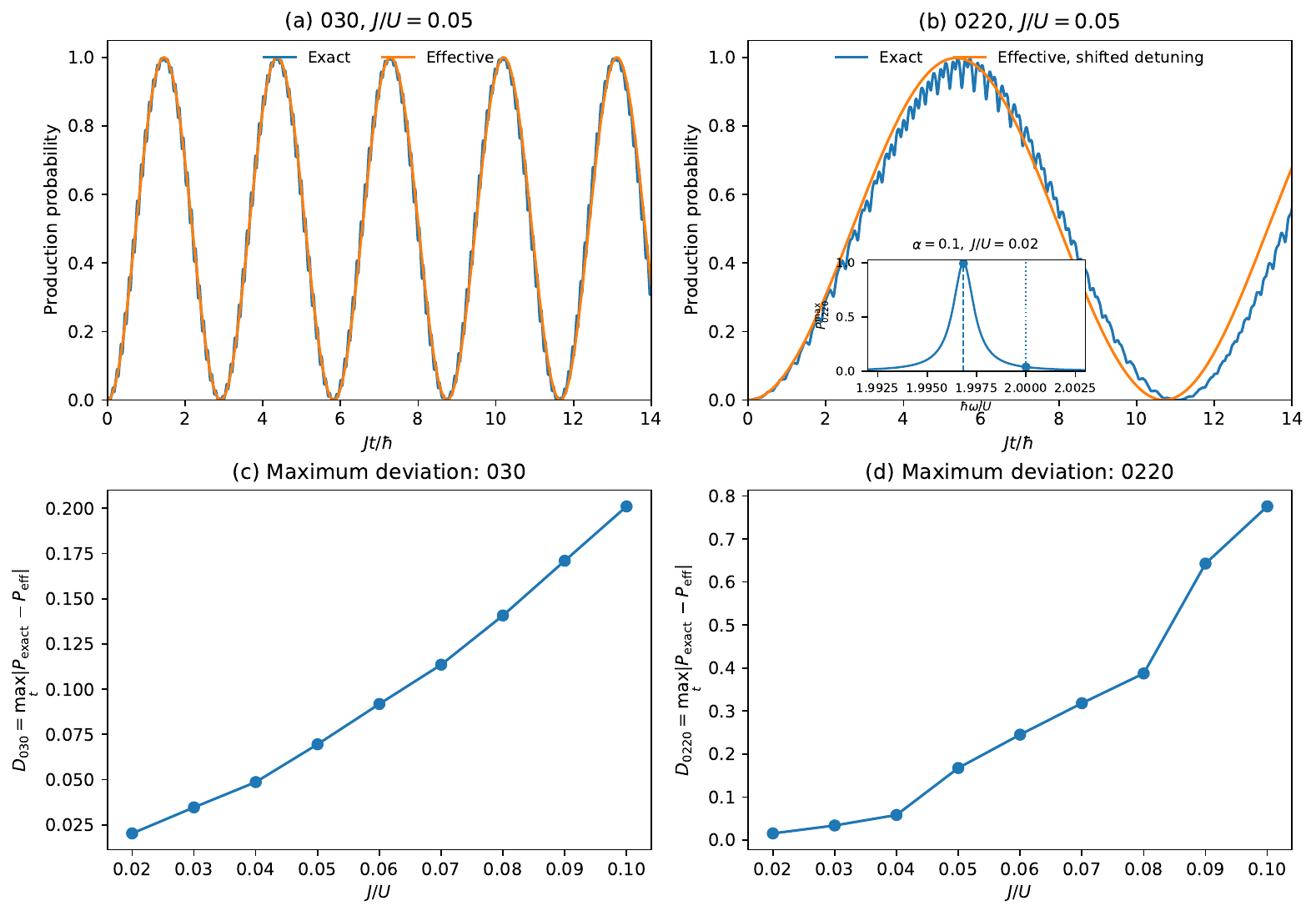}
\caption{Exact benchmark of the local effective descriptions.  Panels (a) and (b) compare exact propagation of the full periodically driven few-site problem with the analytic effective laws for $J/U=0.05$ and $\alpha=1$: direct $030$ nucleation near $\hbar\omega=3U$ in (a), and the second-order $0220$ channel with the self-energy-corrected detuning in (b).  Because $\Sigma_f-\Sigma_i$ is small compared with $G_1$ at $\alpha=1$, the nominal $2U$ frequency remains close to resonance.  The inset of (b) provides the discriminating weak-field test at $J/U=0.02$, $\alpha=0.1$: the continuous curve is the effective maximum-transfer profile, strongly displaced from the atomic value $\hbar\omega/U=2$, while the two markers are exact numerical benchmarks, giving $P_{0220}^{\max}=0.03930$ there and $0.9956$ at the corrected resonance $\hbar\omega/U=1.99684$.  Panels (c) and (d) show the maximum deviation $D=\max_{0\le Jt/\hbar\le14}|P_{\rm exact}-P_{\rm eff}|$ for independently calculated $J/U$ values.  The first-order compact channel is more robust, while the second-order channel becomes increasingly sensitive to higher-order corrections as $J/U$ grows.}
\label{fig:exactvalidation}
\end{figure*}

The weak-field test in the inset of Fig.~\ref{fig:exactvalidation}(b) isolates the role of the diagonal shifts much more sharply.  For
\begin{equation}
J/U=0.02,\qquad \alpha=0.1,
\end{equation}
second-order perturbation theory gives, at the bare atomic resonance, a differential shift close to $-8J^2/U$ while $G_1\propto\alpha J^2/U$.  Solving $\Delta_{1,\mathrm{eff}}^{(0220)}=0$ yields
\begin{equation}
\frac{\hbar\omega_{\rm res}}{U}=1.99684.
\label{eq:weakfieldresnum}
\end{equation}

For this weak-field benchmark, the exact population is maximized over a
frequency-dependent observation window extending to
\begin{equation}
T=1.08\,\frac{\pi\hbar}{2|G_1|},
\end{equation}
with $G_1$ evaluated separately at each drive frequency. This corresponds
to $JT/\hbar\simeq106.35$ at the bare atomic frequency and
$JT/\hbar\simeq106.60$ at the shifted resonance, substantially longer
than the window $0\leq Jt/\hbar\leq14$ used for the $\alpha=1$
comparisons. The exact time evolution was computed with a DOP853
integrator using relative and absolute tolerances of $10^{-9}$ and
$10^{-11}$, respectively, with the weak-field population sampled at
$\Delta(Jt/\hbar)=0.04$. The Bessel sums entering the effective
quantities were truncated at $\ell=\pm30$.

Within these observation windows, the contrast between the bare and
shifted resonance is pronounced. At the bare atomic frequency
$\hbar\omega/U=2$, exact propagation yields only
\begin{equation}
P_{0220}^{\max}=0.03930,
\end{equation}
whereas at the shifted resonance it gives
\begin{equation}
P_{0220}^{\max}=0.9956.
\end{equation}
This discriminating test confirms that the self-energy correction is not
cosmetic: in a weak field it determines whether the system is on or off
resonance. Conversely, the near cancellation at $\alpha=1$ explains why
a nominal $2U$ benchmark can appear accurate even when the diagonal
terms are omitted.  

Figure~\ref{fig:exactvalidation} summarizes both regimes.  The lower panels sample $J/U=0.02,0.03,\ldots,0.10$ independently, with each point obtained from a separate exact propagation.  Both effective descriptions converge toward the exact dynamics in the strong-coupling limit.  The direct channel remains quantitatively more robust because it is first order in $J$, while the $0220$ law relies on eliminating intermediate states and is correspondingly more sensitive to higher-order corrections.  The appropriate conclusion is therefore not that the local formulas are universal Rabi laws for the extended chain, but that they correctly identify the leading microscopic production amplitudes and, once the diagonal shifts are included, the local resonance structure.

\section{Time-resolved production, many-body redistribution, and pair separation}
\label{sec:pairdynamics}

The local benchmark determines whether the few-state effective Hamiltonians reproduce their intended microscopic processes.  The Schwinger interpretation requires a complementary extended-chain question: when the system is driven near the lower pair band, how much weight enters the atomic $2U$ manifold, what other excited sectors are populated, and does the elementary dipole--antidipole component develop a genuine separation between its two dipole centers?

We propagate the full time-dependent problem in Mott-connected open-chain fragments with $n_{\max}=4$, $J/U=0.05$, $\alpha=1$, and
\begin{equation}
\hbar\omega=2U.
\end{equation}
The rotating-frame propagation used numerically is exactly equivalent for occupation probabilities to the laboratory-frame evolution and retains all harmonics; no RWA and no elimination of the triply occupied states is imposed.

Define $\mathcal P_{2U}$ as the projector onto the atomic interaction-energy manifold containing two holes, two doublons, and all other sites singly occupied,
\begin{equation}
P_{2U}(t)=\langle\psi(t)|\mathcal P_{2U}|\psi(t)\rangle.
\label{eq:P2U}
\end{equation}
It is essential to state what this quantity measures: $P_{2U}$ is the probability that the \emph{whole chain} lies in this two-hole/two-doublon atomic manifold.  It is not a density of all produced defects and it does not include configurations at higher atomic interaction energy.

Within the finite observation window
\begin{equation}
0\le Jt/\hbar\le14,
\end{equation}
we therefore define
\begin{equation}
P_{2U}^{\max}(L;T)
=\max_{0\le t\le T}P_{2U}(L,t),
\qquad JT/\hbar=14.
\label{eq:P2maxwindow}
\end{equation}
Starting from the pure Mott state, the maxima are
\begin{equation}
\begin{array}{c|ccc}
L&6&8&10\\ \hline
P_{2U}^{\max}&0.874&0.700&0.330
\end{array},
\label{eq:P2size}
\end{equation}
occurring at $Jt/\hbar\simeq6.20$, $9.03$, and $3.09$, respectively.  The size dependence is real, but it must not be read as a monotonic suppression of total production.  Larger fragments have more pathways into sectors beyond $2U$.

To expose this redistribution, let
\begin{equation}
P_{\ge4U}(t)=
\sum_{E_{\rm at}\ge4U}|\langle\mu|\psi(t)\rangle|^2
\label{eq:Pge4}
\end{equation}
be the total probability in configurations with atomic interaction energy at least $4U$.  At the respective $P_{2U}$ maxima we find
\begin{equation}
P_{\ge4U}(t_*)\simeq
0.0308,\ 0.2820,\ 0.3326
\qquad(L=6,8,10).
\label{eq:Pge4atmax}
\end{equation}
For $L=10$, therefore, the higher-energy population is already essentially as large as $P_{2U}$ at the time when the latter is maximal.  Later in the same window $P_{\ge4U}$ reaches about $0.653$.  This directly shows why $P_{2U}^{\max}$ is not a global production efficiency: population can leave the Mott state, pass through or coexist with the $2U$ manifold, and occupy more complex many-body configurations.

\begin{figure*}[!t]
\centering
\includegraphics[width=1\linewidth]{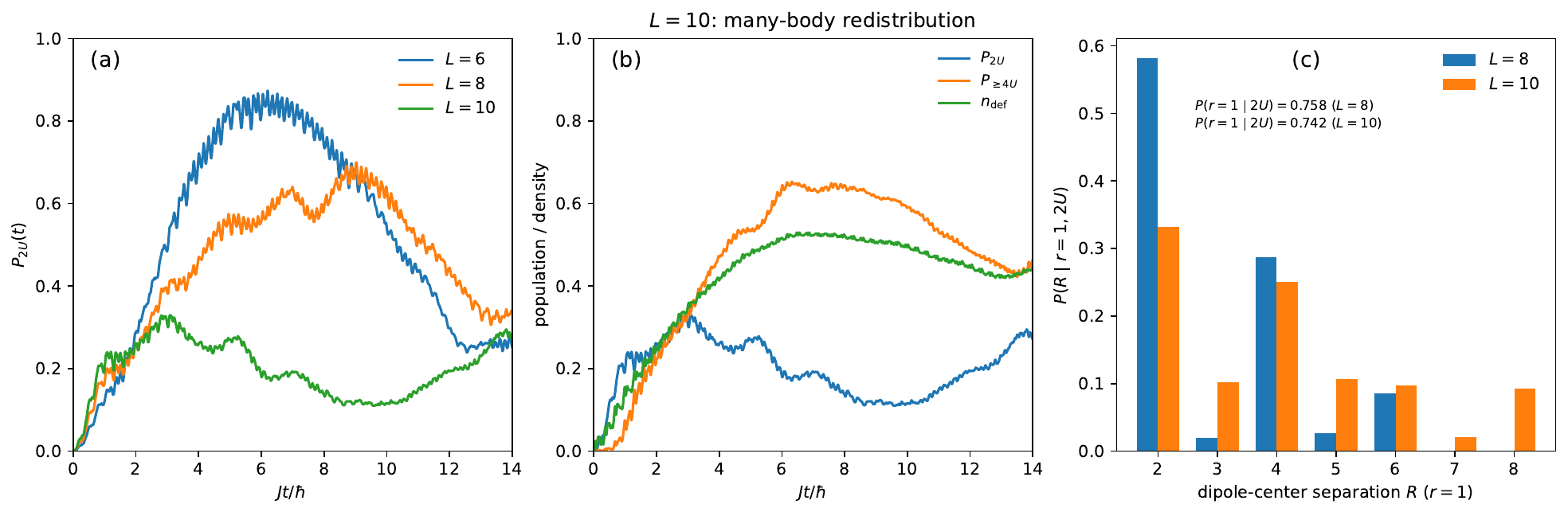}
\caption{Extended-chain production, many-body redistribution, and geometric separation for open chains with $n_{\max}=4$, $J/U=0.05$, $\alpha=1$, and $\hbar\omega=2U$.  (a) Exact time evolution of the two-hole/two-doublon atomic-manifold probability $P_{2U}(t)$ for $L=6,8,10$ within $0\le Jt/\hbar\le14$.  (b) For $L=10$, comparison of $P_{2U}(t)$ with the probability $P_{\ge4U}(t)$ in higher atomic-energy sectors and the defect density $n_{\rm def}(t)$.  The sizable transfer to $E_{\rm at}\ge4U$ shows that a falling $P_{2U}$ does not imply suppressed total excitation production.  (c) Conditional distribution of the dipole-center separation $R$ within the elementary $r=1$ component of the $2U$ manifold, evaluated at each chain's $P_{2U}$ maximum for $L=8$ and $L=10$.  The corresponding elementary-pair fractions are
$P(r=1|2U)=0.758$ and $0.742$. The $L=6$ distribution is omitted from panel (c) because its short
accessible range of $R$ provides little additional geometrical information.
The finite and increasing support at larger $R$ demonstrates spatial
separation without confusing it with internal dipole extension.}
\label{fig:pairdynamics}
\end{figure*}

A complementary local measure is the defect density
\begin{equation}
n_{\rm def}(t)=\frac1L\sum_j
\left\langle1-\Pi_{n_j=1}\right\rangle,
\label{eq:ndef}
\end{equation}
with analogous doublon, hole, and triple-or-higher densities obtainable from occupation-resolved projectors.  Panel (b) of Fig.~\ref{fig:pairdynamics} shows $n_{\rm def}(t)$ together with $P_{2U}(t)$ and $P_{\ge4U}(t)$ for $L=10$.  The three curves separate two notions that would otherwise be conflated: probability of remaining in the $2U$ two-hole/two-doublon manifold and total local departure from the Mott configuration.

As a control against a preparation-quench contribution, we repeated the $L=8$ evolution without the drive and also starting from the exact dressed ground state $|G\rangle$ of $H_0$.  For $A=0$, $P_{2U}$ never exceeds $3.4\times10^{-3}$ over the same window.  With resonant driving from the dressed ground state, it reaches about $0.709$.  The large population of the lower manifold is therefore field induced rather than an artifact of starting from the atomic product state.

We now resolve the geometry inside the $2U$ sector.  For every contributing configuration, the two holes and two doublons are ordered as $h_1<x_1<x_2<h_2$, and dipole conservation enforces the common internal size
\begin{equation}
r=x_1-h_1=h_2-x_2.
\end{equation}
The doublon distance $d=x_2-x_1$ and the dipole-center separation $R$ satisfy $R=d+r$.  For elementary dipoles ($r=1$), the two distances differ by one lattice spacing; for internally extended dipoles, their difference is the internal size $r$.  The conditional joint distribution may be written as
\begin{equation}
P(r,R\,|\,2U;t)=
\frac{\sum_{\mu\in2U}\delta_{r,r_\mu}\delta_{R,R_\mu}
|\langle\mu|\psi(t)\rangle|^2}
{P_{2U}(t)}.
\label{eq:jointsep}
\end{equation}
This definition makes explicit that two different processes are possible in principle: internal extension of each dipole and separation of the two dipoles from one another.

The elementary $r=1$ component remains dominant but not exclusive.  At the time $t_*$ of maximal $P_{2U}$,
\begin{equation}
P(r=1\,|\,2U;t_*)\simeq
0.676,\ 0.758,\ 0.742
\qquad(L=6,8,10).
\label{eq:r1fractions}
\end{equation}
Thus roughly one quarter of the $L=8$ and $L=10$ $2U$ weight lies in internally extended $r>1$ configurations.  Restricting to the elementary $r=1$ family removes the geometric ambiguity and gives a clean Schwinger-type separation observable,
\begin{equation}
P(R\,|\,r=1,2U;t).
\label{eq:conditionalSep}
\end{equation}
For $r=1$, $R=d+1$, so the center-to-center and doublon distances differ only by this fixed offset.  At the respective $2U$ maxima the mean center separation is approximately
\begin{equation}
\langle R\rangle_{r=1}
\simeq2.747,\ 3.019,\ 3.959
\qquad(L=6,8,10),
\label{eq:Rsize}
\end{equation}
corresponding to mean doublon distances $1.747$, $2.019$, and $2.959$.  The larger accessible chains support weight at greater center-to-center separations even after the internally extended $r>1$ configurations are removed from the analysis.  Because the distributions are evaluated at size-dependent times $t^*$, however, the increase of $\langle R\rangle$ with $L$ should not be interpreted as a finite-size scaling law for a propagation velocity.

This separation is qualitatively related to the light-cone spreading of dipole correlations after Mott-to-Mott quenches reported by Oh, Han, and Lee~\cite{OhEtAl2024}.  In contrast, our periodic tensor-field protocol resolves the conditional center-to-center separation of the produced dipole--antidipole pair at fixed internal size $r=1$, without extracting a correlation-front velocity.

The real-space result is therefore more precise than a statement based on doublon distance alone.  The produced $2U$ sector contains both elementary and internally extended dipoles, but the elementary $r=1$ component by itself develops substantial center-to-center separation.  This is fully compatible with exact dipole conservation: the mobile objects are oppositely oriented dipoles, not isolated mobile fracton charges.

\section{Dynamical Schwinger interpretation and experimental perspective}
\label{sec:discussion}

\subsection{Precise Schwinger dictionary}

The extended-chain analysis permits a sharper correspondence than the purely local $3U$ picture:
\begin{equation}
\begin{array}{rcl}
\text{QED} & \leftrightarrow & \text{dipole-conserving lattice}\\
\text{vacuum} & \leftrightarrow & \text{Mott background}\\
E(t) & \leftrightarrow & E_{xx}(t)\\
2mc^2 & \leftrightarrow & E_{\rm edge}(L)\\
e^-e^+ & \leftrightarrow & d\bar d\\
\text{pair separation} & \leftrightarrow & \text{center separation }R\\
\text{multiphoton production} & \leftrightarrow & \text{Floquet }d\bar d\text{ production}.
\end{array}
\label{eq:dictionary}
\end{equation}
The correspondence is operational, not an identity of field theories; the entries in Eq.~\eqref{eq:dictionary} summarize results derived in Secs.~\ref{sec:channel} and \ref{sec:EDspectrum} rather than introducing a second interpretation of them.

Our use of ``Schwinger'' refers specifically to the \emph{multiphoton regime of dynamical Schwinger production}, where time-dependent fields can contain perturbative multiphoton contributions in addition to the nonperturbative constant-field mechanism \cite{GelisTanji2016}.  We have not derived the nonperturbative constant-field exponent $\exp[-\pi m^2/(qE)]$ or its fractonic analogue.  Equation~\eqref{eq:linearSeparationEnergy} suggests an interesting static-field tunneling problem because the tensor-field energy contains the work term $-ArR$ and is linear in the center separation $R$ at fixed internal dipole size $r$.  Pursuing the associated WKB/instanton problem would be a separate project.

\subsection{Experimental observables}

The underlying dipole-conserving dynamics and synthetic tensor drive are already motivated by cold-atom proposals and experiments \cite{LakeEtAl2023,BoeslEtAl2024,KimEtAl2025,ZhangLvZhou2025,ZhangZhang2026}.  The present protocol differs from the manipulation setting of Ref.~\cite{ZhangZhang2026} because the initial state contains no prepared dipole or fracton excitation.

Occupation resolution is essential.  The compact channel contains a triply occupied site, so parity-only fluorescence would not uniquely distinguish $n=1$ from $n=3$.  A local compact-state projector is
\begin{equation}
\hat\Pi_j^{(030)}
=|0,3,0\rangle_{j-1,j,j+1}\langle0,3,0|.
\end{equation}
The first elementary separated configuration can similarly be detected with
\begin{equation}
\hat\Pi_j^{(0220)}
=|0,2,2,0\rangle_{j-1,j,j+1,j+2}\langle0,2,2,0|.
\end{equation}
For the full $2U$ sector, however, the numerical observable and the experimental reconstruction must use the same geometry.  Number-resolved snapshots identify the two holes $h_{1,2}$ and doublons $x_{1,2}$, from which one reconstructs
\begin{equation}
r=x_1-h_1=h_2-x_2,
\qquad
R=(x_2-x_1)+r.
\end{equation}
Projectors onto $|02\,1\cdots1\,20\rangle$ therefore measure specifically the elementary $r=1$ component, while more general $2U$ snapshots also resolve the internally extended $r>1$ population.  This distinction is necessary if an experimental separation distribution is to be compared directly with Eq.~\eqref{eq:jointsep} or Fig.~\ref{fig:pairdynamics}(c).

The many-body redistribution in Fig.~\ref{fig:pairdynamics}(b) suggests additional simple observables.  The total probability in the $2U$ sector can be distinguished from the doublon density, hole density, triple-or-higher occupancy, and the total defect density $n_{\rm def}$.  These quantities determine whether population remains in the $2U$ two-hole/two-doublon manifold or is transferred into more highly excited configurations.

A weak-field frequency scan should reveal qualitatively different spectral structures: a broad lower band derived from $2U$ and a narrower compact manifold near $3U$.  The laboratory-frame response is governed by $S_{Q_2}(E)$, equivalently by the odd-harmonic spectrum $S_{\mathcal O-\mathcal O^\dagger}(E)$ through Eq.~\eqref{eq:spectralidentity}.  The direct lower-band spectral weight is suppressed as $(J/U)^2$ at strong coupling, yet resonant finite-amplitude driving can build a large lower-manifold population and subsequently populate higher sectors, as the exact dynamics demonstrates.

\subsection{Three-body physics and regime of validity}

Triple occupancy introduces both an opportunity and a limitation.  The following three-body term is introduced only to discuss experimental corrections and is not included in the Hamiltonian used in the ED or exact-time calculations above.  If the effective Hamiltonian contains a three-body interaction
\begin{equation}
H_3=\frac W6\sum_jn_j(n_j-1)(n_j-2),
\end{equation}
then
\begin{equation}
\Delta E_{030}=3U+W,
\end{equation}
whereas states built only from doublons remain at $2U$ in the atomic limit.  A repulsive $W$ therefore shifts the compact resonance upward relative to the separated-pair band.  Inelastic three-body loss can also limit the observation time of coherent $030$ dynamics.

The theory is intended for the strong-coupling Mott side.  All explicit calculations reported here use $J/U\leq0.10$, deliberately restricting the analysis to the Mott regime and away from the dipole-condensed phases discussed in Refs.~\cite{LakeHermeleSenthil2022,LakeEtAl2023}.  We do not attempt to locate or approach that transition; the present strong-coupling expansion and few-state effective descriptions should not be extrapolated into the dipole-condensed regime.  There are two independent validity questions.  First, the Floquet RWA requires discarded harmonics to remain off resonant, as quantified by Eq.~\eqref{eq:RWAparameter}.  Second, the local few-state truncations require leakage into the rest of the connected fragment to remain small on the timescale of interest.  Exact diagonalization shows explicitly that these are not the same criterion: one may have small Floquet ratios and still have several-percent lower-band spectral weight.  The extended-chain analysis therefore provides an independent test of many-body leakage beyond the local Floquet criterion.

Finally, the connected fragment depends on the prepared initial configuration.  A vacancy or other preparation defect can change the dynamically accessible fragment even at the same global $(N,P)$.  This is an experimentally relevant consequence of Hilbert-space fragmentation, although a quantitative defect study lies outside the present scope.

\section{Conclusions and outlook}

In summary, a periodic tensor field can produce spatially separating dipole--antidipole excitations from a Mott background even under exact dipole conservation.  The microscopic process displays an unusual separation between energetic and dynamical accessibility: the lowest finite-chain pair sector evolves from the atomic $2U$ manifold, whereas the strongest direct nucleation channel is the compact $3U$ configuration.  Access to the lower band is therefore energetically favored but perturbatively weaker, with its normalized odd-harmonic spectral weight scaling as $(J/U)^2$ at strong coupling.

A consistent second-order treatment further shows that virtual intermediate states shift the lower-channel resonance.  This correction is nearly innocuous for the representative $\alpha=1$ drive, where the endpoint self-energies largely cancel, but becomes decisive in the weak-field regime because the effective coupling narrows linearly with the drive amplitude while the differential shift does not.  Exact few-site propagation confirms that tuning to the self-energy-shifted resonance restores nearly complete local conversion where the bare atomic frequency can leave the system strongly off resonance.

Exact many-body evolution complements this local picture.  The lower $2U$-derived sector contains both elementary and internally extended dipoles, yet the elementary $r=1$ component alone develops substantial center-to-center separation.  At the same time, population is redistributed into higher atomic-energy sectors, so the probability $P_{2U}$ cannot be identified with a total production efficiency.  The no-drive and dressed-ground-state controls confirm that the observed lower-manifold occupation is field induced rather than a preparation artifact.

The resulting correspondence is specifically with the multiphoton regime of dynamical Schwinger production.  Exact dipole conservation does not eliminate field-induced pair creation; it changes the identity of the mobile products from opposite charges to oppositely oriented dipoles, while isolated fractonic charges remain constrained.  We have not derived a nonperturbative constant-field Schwinger exponent or a higher-rank instanton law.  A natural next question is whether a static or slowly varying rank-two field can support a genuinely nonperturbative tunneling regime associated with the field work required to separate opposite dipoles.  The periodically driven results established here provide a controlled starting point for that problem.

\appendix
\section{Floquet--Sambe derivation of the local second-order effective theory}
\label{app:sambe}

For completeness, we derive the local second-order amplitudes used in Sec.~\ref{sec:hierarchy} directly in the extended Floquet (Sambe) space. Let $|i,0\rangle$ denote the initial state $|1111\rangle$ in the reference Floquet sector, $|f,-m\rangle$ the final state $|0220\rangle$ shifted by $m$ drive quanta, and $|a,-\ell\rangle$ the two intermediate states $|0301\rangle$ and $|1030\rangle$ in Floquet sector $-\ell$. Their unperturbed quasienergies are
\begin{align}
\varepsilon_i^{(0)}&=0,\\
\varepsilon_f^{(0)}&=2U-m\hbar\omega,\\
\varepsilon_a^{(0)}&=3U-\ell\hbar\omega.
\end{align}
The $\ell$th Fourier component of the rotating-frame hopping connects neighboring Floquet sectors with matrix element $-J\mathcal J_\ell(\alpha)$ times the corresponding bosonic factor. Each reflected path carries a factor $\sqrt6$ on both hops, and the two path amplitudes add coherently.

Near the $m$-quantum resonance, where the initial and final quasienergies are degenerate to the order retained, standard second-order degenerate perturbation theory gives
\begin{equation}
G_m=-12J^2\sum_{\ell\in\mathbb Z}
\frac{\mathcal J_\ell(\alpha)\mathcal J_{m-\ell}(\alpha)}{3U-\ell\hbar\omega},
\label{eq:appGm}
\end{equation}
which is Eq.~\eqref{eq:M0220} of the main text. Equation~\eqref{eq:appGm} applies provided the intermediate Floquet replicas remain off resonance. If a denominator $3U-\ell\hbar\omega$ vanishes or becomes comparable to the retained couplings, the corresponding replica belongs to
the near-degenerate subspace and must not be eliminated. The same restriction
applies to the diagonal denominators entering Eqs. (A5) and (A6). The
$m=1$ benchmarks of the main text satisfy this condition.

The diagonal corrections follow from virtual excursions of each endpoint into the same intermediate manifold. For the initial state, the denominator is $0-(3U-\ell\hbar\omega)$. For the final state in Floquet sector $-m$, reindexing the exchanged Floquet quanta gives the denominator $-(U-\ell\hbar\omega)$. Summing the two reflected intermediate states yields
\begin{align}
\Sigma_i&=-12J^2\sum_{\ell\in\mathbb Z}
\frac{\mathcal J_\ell^2(\alpha)}{3U-\ell\hbar\omega},
\label{eq:appSigmai}\\
\Sigma_f&=-12J^2\sum_{\ell\in\mathbb Z}
\frac{\mathcal J_\ell^2(\alpha)}{U-\ell\hbar\omega}.
\label{eq:appSigmaf}
\end{align}
These are Eqs.~\eqref{eq:Sigmai} and \eqref{eq:Sigmaf}. Hence
\begin{equation}
\Delta^{(0220)}_{m,\mathrm{eff}}=2U-m\hbar\omega+\Sigma_f-\Sigma_i,
\end{equation}
showing explicitly that the off-diagonal coupling and the differential self-energy shift arise at the same perturbative order and must be retained together in a consistent second-order treatment.

The overall sign follows from
$\varepsilon_i^{(0)}-\varepsilon_a^{(0)}
=-(3U-\ell\hbar\omega)$; it may be reversed by a phase
redefinition of the final basis state and therefore does not affect the
transition probabilities, which depend on $|G_m|$.

\section*{Acknowledgments}
C.R.M. acknowledges partial financial support from the Conselho Nacional de Desenvolvimento Cient\'ifico e Tecnol\'ogico (CNPq), Brazil, through Grant No.~301122/2025-3.
R.N.C.F. acknowledges financial support from CNPq, Brazil, through Grant No.~309908/2022-1.

\bibliographystyle{apsrev4-2}
\bibliography{ref_PRB_updated_DOI}

\end{document}